\documentclass[runningheads]{llncs}
\usepackage[T1]{fontenc}
\usepackage{tikz}
\usetikzlibrary{positioning, shapes.geometric}
\definecolor{blizzardblue}{rgb}{0.67, 0.9, 0.93}
\definecolor{pastelorange}{rgb}{1.0, 0.7, 0.28}
\definecolor{celadon}{rgb}{0.67, 0.88, 0.69}

\usepackage{caption}
\usepackage{bbding}
\usepackage{enumitem}
\usepackage[inference]{semantic}
\usepackage{amsmath, amssymb, stmaryrd}
\usepackage{booktabs}
\usepackage{url}
\usepackage{xcolor}
\usepackage{mathtools}
\usepackage{threeparttable}
\usepackage{proof}
\usepackage{mathpartir}
\usepackage{centernot}
\usepackage{subcaption}
\usepackage{graphicx}
\usepackage{cite}
\usepackage{latexsym}
\usepackage{tabularx}
\usepackage{tabularray}
\UseTblrLibrary{booktabs}
\spnewtheorem*{theoremstar}{Theorem}{\bfseries\upshape}{\itshape}
\usepackage{hyperref}

\newcommand{\shortus}{%
  \leavevmode%
  \kern0.06em%
  \raisebox{-0.5ex}{%
    \rule{0.35em}{0.5pt}%
  }%
  \kern0.06em%
}

\newcommand\wand{\mathrel{-\mkern-6mu*}}

\newcommand{\field}[2]{\mathsf{field\shortus addr}(#1, #2)}
\newcommand{\store}[2]{\mathsf{data\shortus at}(#1, #2)}
\newcommand{\dsl}{Stellis}

\newcommand{\Conv}{\mathop{\scalebox{2.6}{\raisebox{-0.3ex}{$\ast$}}}}

\renewcommand{\vec}[1]{\bar{#1}}

\usepackage{listings}
\usepackage{color,appendix}
\usepackage{multirow}
\usepackage{multicol, cuted}

\makeatletter
\newcommand{\printfnsymbol}[1]{%
  \textsuperscript{\@fnsymbol{#1}}%
}
\makeatother

\begin{document}

\title{\dsl{}: A Strategy Language for Purifying Separation Logic Entailments}
\author{
Zhiyi Wang\inst{1,}\thanks{Both authors contributed equally to this work.}\orcidID{0009-0004-0256-1714}\and
Xiwei Wu\inst{2,}\printfnsymbol{1}\orcidID{0009-0006-2469-3800}\and 
Yi Fang\inst{1}\orcidID{0009-0007-6975-9719}\and
Chengtao Li\inst{1}\orcidID{0009-0009-7358-3941}\and
Hongyi Zhong\inst{2}\orcidID{0009-0001-4994-9159}\and
Lihan Xie\inst{2}\orcidID{0009-0009-5315-5691}\and
Qinxiang Cao\inst{2(}\Envelope\inst{)}\orcidID{0000-0002-5678-6538}\and
Zhenjiang Hu\inst{1}\orcidID{0000-0002-9034-205X}
}
\institute{
Peking University, China\\\email{\{zhiyi.wang, huzj\}@pku.edu.cn\\\{chnfy, lipee\}@stu.pku.edu.cn} \and
Shanghai Jiao Tong University, China\\\email{\{yashen, zhonghongyi1204, sheringham, caoqinxiang\}@sjtu.edu.cn}
}

\authorrunning{Z. Wang et al.}

\maketitle
\begin{abstract}
Automatically proving separation logic entailments is a fundamental challenge in verification. While rule-based methods rely on separation logic rules (lemmas) for automation, these rule statements are insufficient for describing automation strategies, which usually involve the alignment and elimination of corresponding memory layouts in specific scenarios. To overcome this limitation, we propose \textbf{\dsl}, a strategy language for \emph{purifying} separation logic entailments, i.e., removing all spatial formulas to reduce the entailment to a simpler pure entailment. \dsl{} features a powerful matching mechanism and a flexible action description, enabling the \emph{straightforward} encoding of a wide range of strategies. To ensure strategy soundness, we introduce an algorithm that generates a soundness condition for each strategy, thereby reducing the soundness of each strategy to the correctness of its soundness condition. Furthermore, based on a mechanized reduction soundness theorem, our prototype implementation generates correctness proofs for the overall automation. We evaluate our system on a benchmark of 229 entailments collected from verification of standard linked data structures and the memory module of a microkernel, and the evaluation results demonstrate that, with such flexibility and convenience provided, our system is also highly \emph{effective}, which automatically purifies 95.6\% (219 out of 229) of the entailments using 5 libraries with 98 strategies.
\end{abstract}
\section{Introduction}\label{sec:intro} 

Over the past two decades, proof automation in separation logic 
has demonstrated significant power in efficiently verifying memory-related program properties. Numerous verifiers have been developed, including 
VeriFast \cite{jacobs2011verifast}, 
Bedrock \cite{chlipala2011mostly},
HIP/SLEEK \cite{chin2012automated}, 
GrassHopper \cite{piskac2014grasshopper},
Viper \cite{muller2016viper}, 
RefinedC \cite{sammler2021refinedc}, 
Diaframe \cite{mulder2022diaframe}, 
CN \cite{pulte2023cn}, 
and VST-A \cite{zhou2024vst}.

One fundamental challenge in verification is proving that a separation logic entailment holds.
A common approach is to employ a set of separation logic rules (lemmas) of the form $lhs \vdash rhs$ to solve the entailment. For the entailment 
$
\mbox{\it antecedent} \vdash \mbox{\it consequent}, 
$
if $lhs$ matches with the antecedent or $rhs$ matches with the consequent, then the rule can be applied to change the antecedent or the consequent, respectively. The rule-based methods~\cite{berdine2006smallfoot, nguyen2008enhancing} for automating this approach involve iteratively searching for applicable rules, and applying them to simplify the entailment toward a trivial goal (e.g., a tautology).

However, rule statements are inconvenient for describing automation strategies, which usually involve the alignment and elimination of corresponding memory layouts---a ubiquitous transformation in solving separation logic entailments. For example, consider the following entailment encountered in proving functional correctness of a linked list:
\[
\mathsf{lseg}(x, y, l_1) * H ~\vdash~ \exists~l_2, \mathsf{listrep}(x, l_2) * H'
\]
where $\mathsf{lseg}(x, y, l_1)$ indicates a linked list segment from $x$ to $y$, with elements described by the functional list $l_1$; $\mathsf{listrep}(x, l_2)$ indicates a complete linked list from $x$, with elements described by $l_2$; $H$ and $H'$ abbreviate other conjuncts.
We may conclude that the memory represented by $\mathsf{lseg}(x, y, l_1)$ is a part of the memory represented by $\mathsf{listrep}(x, l_2)$ (since both memory start from the same address $x$), and desire an automation strategy: 
\begin{quote}
\emph{
Eliminate the memory of $\mathsf{lseg}(x, y, l_1)$ in the antecedent with part of the memory of $\mathsf{listrep}(x, l_2)$ in the consequent, leaving a portion represented by $\mathsf{listrep}(y, l_3)$, and a pure constraint stating that the concatenation of $l_1$ and $l_3$ should be equal to $l_2$.
}
\end{quote}
With this strategy, we may simplify the above entailment to
\[
H ~\vdash~ \exists~l_2~l_3, l_2==\mathsf{app}(l_1, l_3)\land \mathsf{listrep}(y, l_3) * H'
\]
where $\mathsf{app}$ denotes the list-append function. Repeatedly applying other similar strategies will simplify this entailment further, and ultimately \emph{purify} this entailment by eliminating all spatial formulas that describe memory layouts and reducing it to a simpler pure entailment.

To implement the above strategy, one can theoretically use the following rule to split the \textsf{listrep} predicate in the consequent:
\[
\mathsf{lseg}(p, q, l_1) * \mathsf{listrep}(q, l_2) \vdash \mathsf{listrep}(p, \mathsf{app}(l_1, l_2))
\]
However, this represents an indirect way of implementing strategies, which requires users to first identify the correct rule (or rule combinations)--a ubiquitous but not-so-straightforward task. Can we \emph{directly describe} strategies instead of identifying the correct rule combinations? If so, how can we ensure that our strategy description is \emph{sound/correct}?

In this paper, we propose \textbf{\dsl}, a lightweight domain-specific strategy language for purifying separation logic entailments. To facilitate straightforward encoding of strategies, \dsl{} offers (1) a powerful matching mechanism to identify applicable scenarios via pattern matching, and (2) a flexible action description to describe the desired transformation through adding/removing formulas and introducing auxiliary variables. To ensure soundness, we propose an algorithm that takes a \dsl{} strategy as input and generates its soundness condition as output, and mechanize a reduction soundness theorem in Rocq, which together guarantee that if the soundness condition is proved correct by users, then the strategy is sound. Furthermore, our implemented system generates a complete correctness proof in Rocq for the overall purification process. This endows our system with the \emph{foundational property}: the implementation itself does not need to be trusted, as its output--a formal proof--can be independently verified by Rocq's type checker.

The main technical contributions of the paper can be summarized as follows.  
\begin{itemize}  
    \item We design \dsl{}, a strategy language for purifying separation logic entailments. \dsl{} allows users to declaratively define when, how, and in what order a set of actions is applied to manipulate entailments. This enables the construction and reuse of useful strategy libraries. We present the formal definition of its syntax and semantics, together with some illustrative examples to demonstrate its expressiveness.
      
    \item We provide a systematic method to \emph{guarantee the soundness} of strategies defined in \dsl{}. More specifically, we present an algorithm for generating soundness conditions for user-defined strategies, mechanize a reduction soundness theorem in Rocq, and our implemented system will generate correctness proofs for the simplification of input entailments.
      
    \item We show the details of our implemented system and evaluate it on a benchmark consisting of 229 general entailments generated from the verification of standard linked data structures and the memory module of a microkernel. The results demonstrate the effectiveness of our system, which \emph{automatically purify 95.6\% (219 out of 229) of the entailments} using 5 libraries with 98 strategies.
\end{itemize}
  
\section{Overview}\label{sec:overview}

In this section, we give an overview of our strategy language--\dsl{}, with three concrete examples to show how we can describe automation strategies in \dsl{}.

\subsection{\dsl{} by Example} \label{sec:overview:1}

\dsl{} provides flexible patterns and intuitive actions for writing automation strategies. For example, the strategy S0 in Figure~\ref{fig:overview-example} describes the strategy mentioned in the introduction. Lines~3-4 show the \emph{pattern conditions}, which specify the scenario when strategy S0 should be applied, i.e., when a formula of the form \lstinline|lseg(p, q, l1)| appears in the antecedent and a formula of the form \lstinline|listrep(p, l2)| appears in the consequent. Lines~6-10 then show a sequence of operations describing how to simplify the entailment, that is, first removing the matched formulas (Lines~6-7), then introducing a fresh existential variable (Line~8), and finally adding back two new formulas (Lines~9-10). Hence, to write an automation strategy in \dsl{}, users need only: (1) describe the application scenario through pattern conditions, and (2) specify the desired effects through a sequence of operations.

While a curious reader may immediately wonder about the soundness of Strategy S0--a topic we detail in Section~\ref{sec:overview:2}--let us first explore how \dsl{} facilitates conservative strategy application through additional pattern conditions. For example, sometimes we want to eliminate two \textsf{lseg} predicates that represent the same memory. Strategy S1 in Figure~\ref{fig:overview-example} does this by identifying the \textsf{lseg} predicates (Lines~3-4), removing them (Lines~7-8) and leaving a pure constraint (Line~9) stating that the elements within both memory regions must be equal. Importantly, we constrain this strategy to be applied only if an additional consecutive \textsf{listrep} predicate is observed in the consequent (Line~5). This constraint is necessary, because direct elimination of \textsf{lseg} predicate can sometimes be overly aggressive and transform a provable entailment into an unprovable one\footnote{Consider the entailment $\mathsf{lseg}(p, q, l_1) * \mathsf{lseg}(q, q, l_2) \vdash \exists~l_3, \mathsf{lseg}(p, q, l_3)$. Direct elimination of the \textsf{lseg} predicate transforms this entailment into $\mathsf{lseg}(q, q, l_2)\vdash \exists~l_3, \mathsf{emp}\land l_3 == l_1$, which is unprovable and indicates a memory-leak.}. The presence of a consecutive \textsf{listrep} predicate rules out the potential cycle that could be hidden inside both \textsf{lseg} predicates, leading to a more conservative but safe elimination.




\begin{figure}[t]
    \centering
    \footnotesize
    \ttfamily
    \begin{minipage}[t]{0.48\linewidth}
    \vspace{0pt}
\begin{lstlisting}[numbers=left, xleftmargin=2em, basicstyle=\ttfamily\footnotesize]
// Strategy S0
priority: 1
left:   lseg(?p, ?q, ?l1)
right:  listrep(p, ?l2)
action:
  left_erase(lseg(p, q, l1));
  right_erase(listrep(p, l2));
  exist_add(l3);
  right_add(l2 == app(l1, l3));
  right_add(listrep(q, l3));
\end{lstlisting}
    \end{minipage}
    \hfill
    \begin{minipage}[t]{0.48\linewidth}
\vspace{0pt}
\begin{lstlisting}[numbers=left, xleftmargin=2em, basicstyle=\ttfamily\footnotesize]
// Strategy S1
priority: 0
left:   lseg(?p, ?q, ?l1)
right:  lseg(p, q, ?l2)
        listrep(q, ?l3)
action:
  left_erase(lseg(p, q, l1));
  right_erase(lseg(p, q, l2));
  right_add(l2 == l1);
\end{lstlisting}
    \end{minipage}

\begin{minipage}[t]{0.6\linewidth}
        \vspace{0pt}
\begin{lstlisting}[numbers=left, xleftmargin=2em, basicstyle=\ttfamily\footnotesize]
// Strategy S2
priority: 1
left:   listrep(?p, ?l1)
right:  listrep(p, ?l2)
action: left_erase(listrep(p, l1));
        right_erase(listrep(p, l2));
        right_add(l2 == l1);
\end{lstlisting}
    \end{minipage}
    \hfill
    \begin{minipage}[t]{0.36\linewidth}
    \end{minipage}

    \caption{Example strategies in \dsl{}}
    \label{fig:overview-example}
    \vspace{-1em}
\end{figure}

Another important feature is that users can control the application order of strategies by specifying priorities. For example, the application of strategy S0 will remove a \textsf{listrep} predicate from the consequent, and could potentially break the pattern conditions of strategy S1, making it not applicable. We can assign a higher priority to strategy S1 (Line~2) to avoid this circumstance.

Strategy S2 is similar to strategy S1 and removes identical memory regions represented by \textsf{listrep} predicates. These strategies in Figure~\ref{fig:overview-example} form a \emph{strategy library} that can be widely used in automatically purifying entailments within singly-linked-list theory. For example, they can purify the following entailment:

\begin{small}
\begin{subequations}
\begin{align*}
    &~\mathsf{lseg}(p, q, l_1) * \mathsf{lseg}(q, r, l_2) * \mathsf{listrep}(r, l_3)~\vdash~\exists~l_4~l_5, \mathsf{lseg}(p, q, l_4) * \mathsf{listrep}(q, l_5)\\
    \leadsto&~\mathsf{lseg}(q, r, l_2) * \mathsf{listrep}(r, l_3)~\vdash~\exists~l_4~l_5, \mathsf{listrep}(q, l_5)\land l_4 == l_1\tag{apply~S1}\\
    \leadsto&~\mathsf{listrep}(r, l_3)~\vdash~\exists~l_4~l_5~l_6, \mathsf{listrep}(r, l_6)\land l_4 == l_1\land l_5 == \mathsf{app}(l_2, l_6)\tag{apply~S0}\\
    \leadsto&~\mathsf{True}~\vdash~\exists~l_4~l_5~l_6, l_4 == l_1\land l_5 == \mathsf{app}(l_2, l_6)\land l_6 == l_3\tag{apply~S2}
\end{align*}
\end{subequations}
\end{small}

\subsection{Soundness Guarantee}\label{sec:overview:2}
A strategy is considered sound if, whenever its transformation result is provable, the original entailment must also be provable. The expressiveness of \dsl{} carries the potential risk of introducing unsound strategies. For instance, the soundness of S0 is not immediately evident, while the following strategy is clearly unsound:
\begin{lstlisting}[numbers=left, xleftmargin=2em]
right: listrep(?p, ?l)
action: right_erase(listrep(p, l));
\end{lstlisting}

To ensure soundness, we propose an algorithm (detailed in Section~\ref{sec:soundness}) to generate a soundness condition for each strategy such that if the soundness condition is proved, the strategy is sound. For example, the soundness condition for strategy S0 is:
\[
  \mathsf{lseg}(p, q, l_1) * \mathsf{listrep}(q, l_3) \land l_2 == \mathsf{app}(l_1, l_3) ~\vdash~ \mathsf{listrep}(p, l_2)
\]
An observant reader might notice that this condition is equivalent to the rule mentioned in the introduction. Indeed, this highlights a key advantage: our soundness algorithm offloads this task from the user by automatically generating the appropriate rule.


\begin{figure*}[t]
    \centerline{\includegraphics[width=\linewidth]{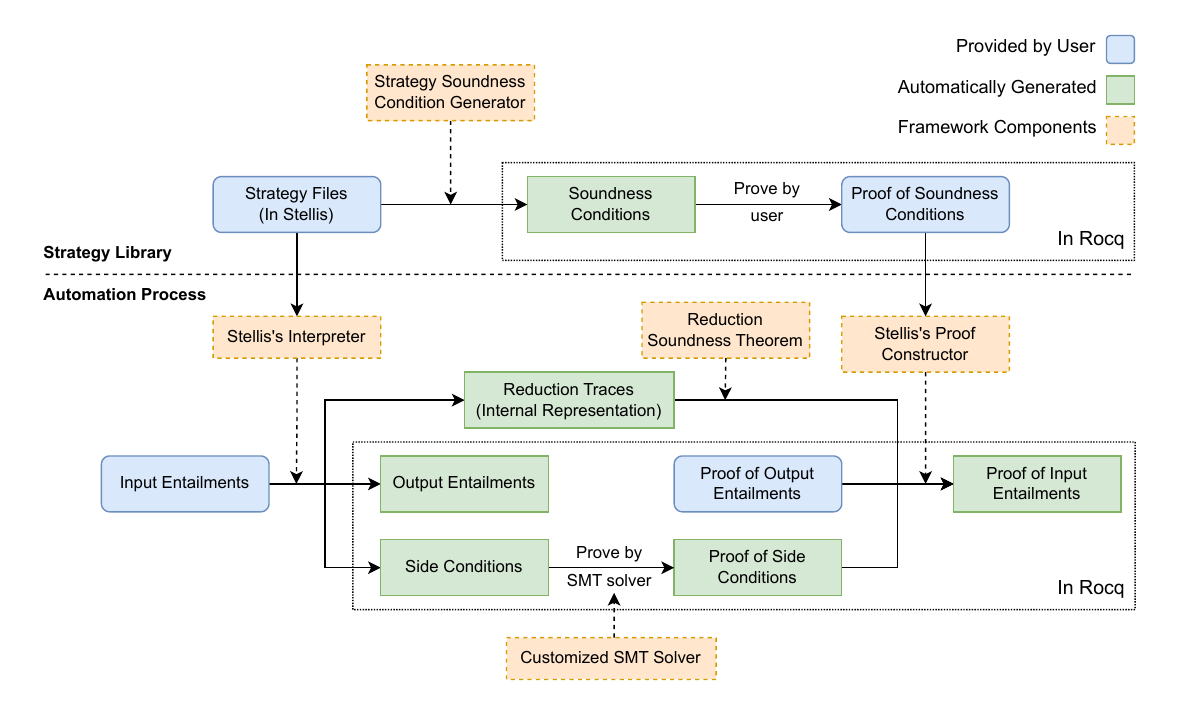}}
    \caption{The workflow of our framework}
    \label{fig:workflow}
    \vspace{-1em}
\end{figure*}

\subsection{Putting Things Together}\label{sec:overview:3}
Figure~\ref{fig:workflow} shows the workflow of our framework. The ideal workflow involves four steps: First, write strategies in \dsl{}; second, invoke the strategy soundness condition generator to generate soundness conditions and prove them to ensure the written strategies are sound; third, invoke \dsl{}'s interpreter, which leverages the written strategies to purify input entailments; Finally, invoke \dsl{}'s proof constructor to generate the Rocq proof for the purification process. A full discussion of the implementation details and the automation process is presented in Section~\ref{sec:implementation}.
\section{Syntax, Semantics and Expressiveness}\label{sec:strategy} 

\subsection{Syntax}\label{sec:strategy:1}

Since strategies are applied to entailments, we first define entailment syntax (detailed in Figure~\ref{fig:entailment-syntax}). Both sides of an entailment are in symbolic-heap~\cite{berdine2005symbolic}, i.e., each side consists of a conjunction of pure formulas and a separating conjunction of spatial formulas. The consequent may be existentially quantified and the whole entailment may be universally quantified. To reason about real-world C programs, we support the term $\field{t}{\text{field}}$ to represent the value of \lstinline|&(t->field)| in C.

Figure~\ref{fig:strategy-syntax} presents the formal syntax of \dsl{}. A \dsl{} program $Prog$ consists of a sequence of strategies $\vec{S}$. A strategy $S$ has the following four elements: (1) A priority number $n$ where smaller numbers indicate higher priority, (2) a sequence of patterns $\vec{q}$, (3) a sequence of checks $\vec{c}$, and (4) an action $a$. For \lstinline|right| patterns, the syntax \lstinline|exists| $x, q_r$ enforces that the pattern variable $x$ binds to an existential variable. The check part ensures the entailment satisfies specific constraints, where \lstinline|left_absent|$(p)$ confirms the absence of a pure formula $p$ in the antecedent, and \lstinline|infer|$(p)$ invokes an SMT solver to determine whether a pure fact $p$ can be inferred from the antecedent. The action part consists of two types: an operation sequence $\vec{o}$ that manipulates the entailment by adding/removing formulas and introducing fresh variables, and \lstinline|instantiate|$(x \to t)$ for instantiation of an existential variable $x$ with a term $t$\footnote{Although instantiation cannot be combined with other operations, this does not limit expressiveness and we show in Appendix~\ref{appendix:strategy-example} how to circumvent the limitation.}.

\begin{figure}[t]
\begin{align*}
&t ::= n \mid x \mid \field{t}{\text{field}} \mid h(t_1, t_2, ...)\\
&p ::= \mathsf{True}\mid t_1 == t_2 \mid {\sim}p \mid p_1 \oplus p_2 \mid P(t_1, t_2, ...)\\
&s ::= \mathsf{emp} \mid \store{t_1}{t_2} \mid A(t_1, t_2, ...)\\
&f ::= p \mid s \qquad n ::= 0 \mid 1 \mid \cdots \qquad \oplus ::= \land \mid \lor \mid \,\to\, \mid
\,\leftrightarrow\\
&E ::= \forall\,\vec{x}_{\forall}, \bigwedge\nolimits_i p_{i} \land
\Conv\nolimits_i s_i \vdash \exists\,\vec{x}_{\exists}, \bigwedge\nolimits_i p_{i} \land
\Conv\nolimits_i s_i
\end{align*}
\vspace{-1.5em}
    \caption{Syntax of entailments. $t$ represents terms, $p$ and $s$ respectively represent pure and spatial formulas, and $E$ represents entailments.}
    \label{fig:entailment-syntax}
    \vspace{-1.5em}
\end{figure}

\begin{figure}[t]
\centering
\begin{align*}
&q_l ::= \hat{f} \qquad q_r ::= \texttt{exists }\, x, q_r\mid \hat{f}\qquad q ::= \texttt{left: }q_l\mid \texttt{right: }q_r \\
&c ::= \texttt{left\_absent}(p) \mid \texttt{right\_absent}(p) \mid \texttt{infer}(p) \\
&o ::= \texttt{left\_add}(f) \mid \texttt{right\_add}(f) \mid \texttt{left\_erase}(f) \mid \texttt{right\_erase}(f)\\
&~~\mid \texttt{forall\_add}(x) \mid \texttt{exist\_add}(x)\\
&a ::= \vec{o} \mid \texttt{instantiate}(x\to t)\\
&S ::= (\texttt{priority: }n;~ \vec{q};~\texttt{check: }\vec{c};~\texttt{action: }a)\\
& Prog ::= \vec{S}
\end{align*}
\vspace{-1em}
    \caption{Syntax of {\dsl}. $\hat{f}$ extends the syntax of $f$ with pattern variables $?x$ for term binding.}
    \label{fig:strategy-syntax}
    \vspace{-1em}
\end{figure}

\subsection{Semantics}\label{sec:strategy:2}

The semantics of a \dsl{} program formalizes the automation process, which attempts to iteratively apply strategies according to their priorities, until no strategy is applicable.
This process by design contains no backtracking: once a strategy is applied, the automation process does not backtrack to consider other applicable choices. This design is based on the empirical observation that most purification processes can be completed using conservative strategies, without requiring backtracking. Our prototype system validates this premise, which successfully purifies most of entailments in the evaluation without any backtracking.

The semantics of a strategy is defined by the process of transforming an entailment $E$ into another entailment $E'$. The process consists of the following phases:
\begin{enumerate}
    \item[(1)] Match the pattern part $\vec{q}$ against the entailment $E$, and compute a pattern substitution $\sigma$ (which maps pattern variables $\mathord{?}x$ to terms $t$);
    \item[(2)] Verify that $E$ satisfies the check conditions $\vec{c}$ under $\sigma$;
    \item[(3)] Perform the action $a$ to $E$ under $\sigma$, resulting in $E'$;
    \item[(4)] Ensure that the transformed entailment $E'$ is \textit{well-formed}, i.e., $E'$ is closed, and the set of universal quantifiers does not overlap with the set of existential quantifiers.
\end{enumerate}
If all phases are successful, then $S$ is applicable for transforming $E$ into $E'$.

\subsection{Strategy Examples and Expressiveness Discussion} \label{sec:strategy:3}

While Section~\ref{sec:overview} has thus far focused on singly-linked lists, \dsl{} is capable of handling much more. In this section, we extend our discussion to arrays, and further demonstrate the use of \dsl{} for frame inference.

\begin{figure}[t]
    \centering
    \tiny
    \ttfamily

\begin{minipage}[t]{0.48\linewidth}
        \vspace{0pt}
\begin{lstlisting}[basicstyle=\ttfamily\scriptsize]
// Strategy S3
left: store_array(?p, ?x, ?y, ?l)
right: data_at(p + 4 * ?i, ?v)
check: infer(x <= i);
       infer(i < y);
action:
  left_erase(store_array(p, x, y, l));
  right_erase(data_at(p + 4 * i, v));
  left_add(store_array_hole(p, x, y, i, l));
  right_add(v == nth(i - x, l));
\end{lstlisting}
    \end{minipage}
    \hfill
    \begin{minipage}[t]{0.5\linewidth}
\begin{lstlisting}[basicstyle=\ttfamily\scriptsize]
// Strategy S4
left: store_array_hole(?p, ?x, ?y, ?i, ?l)
right: store_array(p, x, y, ?l1)
check: infer(x <= i);
       infer(i < y);
action:
  left_erase(store_array_hole(p, x, y, i, l));
  right_erase(store_array(p, x, y, l1));
  exist_add(v);
  right_add(data_at(p + 4 * i, v));
  right_add(l1 == update_nth(i - x, v, l));
\end{lstlisting}
    \end{minipage}
\vspace{-1em}
    \caption{More example strategies in \dsl{}}
    \label{fig:strategy-example}\
    \vspace{-1em}
\end{figure}


\textit{Arrays.} We use the predicate $\mathsf{store\shortus array}(p, x, y, l)$ to denote that the slice of an integer array $p[x:y]$ contains the elements $l$. Similarly, $\mathsf{store\shortus array\shortus hole}(p, x, y,\allowbreak i, l)$ represents the same slice but excluding the ownership of the element at index $i$. Consider a typical array usage pattern where (1) one first isolates the permission of $p[i]$ from $\mathsf{store\shortus array}(p, 0, n, l)$, performs some modification, and (2) subsequently re-establishes the predicate as $\exists~l', \mathsf{store\shortus array}(p, 0, n, l')$.
The first step will generate the following proof obligation, which can be simplified by strategy S3 in Figure~\ref{fig:strategy-example}:
\begin{align*}
&~0 \le i < n\land \mathsf{store\shortus array}(p, 0, n, l) * \cdots~\vdash~ \exists~v, \store{p + 4 * i}{v} * \cdots\\
\leadsto&~ 0 \le i < n \land \mathsf{store\shortus array\shortus hole}(p, 0, n, i, l) * \cdots ~\vdash~ \exists~v, v = \mathsf{nth}(i - 0, l) \land \cdots
\end{align*}
The second step will generate the following proof obligation, which can be simplified by strategy S4 in Figure~\ref{fig:strategy-example}:
\begin{align*}
&~0 \le i < n\land \mathsf{store\shortus array\shortus hole}(p, 0, n, i, l) * \store{p + 4 * i}{v} * \cdots ~\vdash~ \\
&~\quad\exists~l', \mathsf{store\shortus array}(p, 0, n, l') * \cdots\\
\leadsto&~ 0 \le i < n \land \store{p + 4 * i}{v} * \cdots~\vdash~ \\
&~\quad\exists~l'~v', l' = \mathsf{update\shortus nth}(i - 0, v', l) \land \store{p + 4 * i}{v'} * \cdots
\end{align*}

\textit{\dsl{} for frame inference.} Modern verifiers routinely perform frame inference during the symbolic execution (or type-checking) of memory accesses and function calls. For example, suppose the current symbolic state is $P$, and the subsequent command attempts to read from an array cell $p[i]$. This necessitates partitioning the current state into the specific permission for $p[i]$ and a residual frame $\mathord{?}Q$, formalized as:
\[
P ~\vdash~ \exists~v, \store{p + 4 * i}{v} * \mathord{?}Q
\]
The task of finding $\mathord{?}Q$ is known as frame inference. \dsl{} is naturally suited for this task. We can model this problem by directly purifying the entailment $P \vdash \exists~v, \store{p + 4 * i}{v}$. If the purification process yields a residual entailment $P' \vdash \phi_{pure}$ (where $\phi_{pure}$ contains only pure conjuncts), then $P'$ constitutes the inferred frame exactly. In the Appendix~\ref{appendix:program-example}, we provide a complete example where \dsl{} is employed for both frame inference and entailment purification, and successfully purifies all generated proof obligations.
\section{Soundness Guarantee}\label{sec:soundness}
As discussed in Section~\ref{sec:overview:2}, users can use \dsl{} to write unsound strategies. In this section, we present an algorithm for generating a soundness condition for a given strategy, such that the correctness of the condition guarantees the soundness of the strategy.

Let us first define the concept of soundness for a strategy. A strategy $S$ is \textit{sound} if and only if for every application of $S$ that transforms an entailment $E$ to another entailment $E'$, the proposition $\llbracket E'\rrbracket \Rightarrow \llbracket E\rrbracket$ is valid. In other words, if we can obtain the proof of the original entailment $E$ from the proof of the transformed entailment $E'$, then this transformation is obviously sound.

To get an intuition of the algorithm, we first consider the following strategy~$S$:
\begin{lstlisting}[numbers=left, xleftmargin=2em]
// Strategy S
left :   data_at(?p, ?v0)
right :  data_at(p, ?v1)
action : left_erase(data_at(p, v0));
         right_erase(data_at(p, v1));
         right_add(v1 == v0);
\end{lstlisting}
and a specific application of $S$ that transforms $E$ into $E'$:\footnote{In the logical formulas, we abbreviate $\store{p}{v}$ as $p \mapsto v$.}
\begin{subequations}
\begin{align}
~&p\mapsto v * H~\vdash~ \exists\, v_1, p\mapsto v_1 * H'\tag{$E$}\\
\leadsto~&H ~\vdash~ \exists\, v_1, v_1 = v_0 \land H'\tag{$E'$}
\end{align}
\end{subequations}

To prove that $\llbracket E'\rrbracket \Rightarrow \llbracket E\rrbracket$ is valid, the key observation is that, since this transformation is captured by a sequence of operations, we can generate the soundness condition based on these operations. The soundness condition $\phi$ for strategy $S$ is:
\begin{align*}
    \phi~\triangleq~ p\mapsto v_0 \vdash \forall\, v_1, v_1 = v_0 \wand p\mapsto v_1.
\end{align*}
where the conjuncts are adopted from the operations inside $S$. With $\phi$, we can prove $\llbracket E'\rrbracket \Rightarrow \llbracket E\rrbracket$ as follows (here we directly prove $E$ and use $E'$ as an assumption):
\begin{small}
\begin{subequations}
\begin{align*}
    &~E =p\mapsto v_0 * H~\vdash~ \exists\, v_1, p\mapsto v_1 * H'\\\displaybreak[0]
    \Leftarrow&~(\forall\, v_1, v_1 = v_0 \wand p\mapsto v_1) * H~\vdash~\exists\, v_1, p\mapsto v_1 * H'\tag{apply~$\phi$}\\\displaybreak[0]
    \Leftarrow&~(\forall\, v_1, v_1 = v_0 \wand p\mapsto v_1) * (\exists\, v_1, v_1 = v_0 \land H')~\vdash~\exists\, v_1, p\mapsto v_1 * H'\tag{apply~$E'$}\\\displaybreak[0]
    \Leftarrow&~(\forall\, v_1, v_1 = v_0 \wand p\mapsto v_1) * (v_1 = v_0 \land H')~\vdash~\exists\, v_1, p\mapsto v_1 * H'\tag{extract~$v_1$}\\\displaybreak[0]
    \Leftarrow&~v_1 = v_0 \land (\forall\, v_1, v_1 = v_0 \wand p\mapsto v_1) * H'~\vdash~\exists\, v_1, p\mapsto v_1 * H'\tag{rearrange}\\\displaybreak[0]
    \Leftarrow&~p\mapsto v_1 * H'~\vdash~\exists\, v_1, p\mapsto v_1 * H' \tag{def. of $\wand$}\\\displaybreak[0]
    \Leftarrow&~ \mathsf{True}
\end{align*}
\end{subequations}
\end{small}

However, focusing solely on the action part of a strategy while ignoring the information embedded in the pattern and check part may lead to an unprovable soundness condition. The pattern part indicates the existence of certain formulas, while the check part may invoke an SMT solver to determine the validity of certain propositions. Our algorithm also accounts for this information and incorporates it into the generated soundness condition.

In the rest of this section, we will elaborate the key ideas of our algorithm. Due to space limit, the full formal description of our algorithm can be found in Appendix~\ref{appendix:soundness}.

\subsection{Key Ideas} \label{sec:soundness:1}
We start by explaining our key ideas about how to generate a soundness condition based on the operations, and then discuss how to incorporate the information embedded in the pattern and check part into the soundness condition.

\textit{The basic algorithm (without quantifiers).} Suppose the current entailment $E$ is $A * H \vdash B * H'$, and we apply the following operations to transform $E$ to $E' \triangleq C * H \vdash D * H'$:
\begin{lstlisting}[numbers=left, xleftmargin=2em]
left_erase(A); right_erase(B);
left_add(C); right_add(D);
\end{lstlisting}
If $\phi \triangleq A \vdash C * (D \wand B)$ holds (inspired by the \textsc{Ramify} rule~\cite{hobor2013ramifications}), then $\llbracket E'\rrbracket \Rightarrow \llbracket E\rrbracket$ can be proved as follows:
\begin{subequations}
\begin{align*}
    &~E = A * H ~\vdash~ B * H'\\
    \Leftarrow&~C * H * (D\wand B)~\vdash~ B * H'\tag{apply~$\phi$}\\
    \Leftarrow&~D * H' * (D \wand B) ~\vdash~ B * H'\tag{apply~$E'$}\\
    \Leftarrow&~B * H' ~\vdash~ B * H' \Leftarrow \mathsf{True}\tag{def. of $\wand$}
\end{align*}
\end{subequations}
In general, any sequence of operations will have a similar overall effect: removing a set of formulas $A$ (from the antecedent) and $B$ (from the consequent), and adding a set of formulas $C$ (to the antecedent) and $D$ (to the consequent). The resulting soundness condition can be expressed as:
\[
    \phi ~\triangleq~ A\vdash C * (D \wand B)
\]

\textit{Dealing with quantifiers.} Now let's try to involve quantifiers into discussion. Suppose the current entailment $E$ is $\forall~a, A*H\vdash \exists~b,  B*H'$, and we apply the following operations to transform $E$ to $E' \triangleq \forall~a~p, C * H \vdash \exists~b~q, D * H'$:
\begin{lstlisting}[numbers=left, xleftmargin=2em]
left_erase(A); right_erase(B); left_add(C);
right_add(D); forall_add(p); exist_add(q);
\end{lstlisting}
where $a, b, p$ and $q$ may each represent a set of quantifiers; $E$ and $E'$ should be well-formed. This time, the generated condition $\phi$ should be
$$
\phi~\triangleq~ A \vdash \exists~p, C * (\forall~v, D\wand B)
$$
where $v$ collects the variables that \textit{only appear in} the conjuncts $D$ and $B$ (but not in $A, C$ and $p$). Following a similar process, we can also prove that $\llbracket E'\rrbracket \Rightarrow \llbracket E\rrbracket$:
\begin{subequations}
\begin{align*}
    &~E = \forall~a, A * H ~\vdash~ \exists~b, B * H'\\
    \Leftarrow&~\forall~a, (\exists~p, C * (\forall~v, D\wand B)) * H~\vdash~\exists~b,  B * H'\tag{apply~$\phi$}\\
    \Leftarrow&~\forall~a~p, (C * H) * (\forall~v, D\wand B)~\vdash~\exists~b,  B * H'\tag{extract~$p$, rearrange}\\
    \Leftarrow&~\forall~a~p, (\exists~b~q, D * H') * (\forall~v, D\wand B)~\vdash~\exists~b,  B * H'\tag{apply~$E'$}\\
    \Leftarrow&~\forall~a~p~b~q, D * (\forall~v, D\wand B) * H' ~\vdash~\exists~b,  B * H'\tag{extract~$b~q$, rearrange}\\
    \Leftarrow&~\forall~a~p~b~q, D * (D\wand B) * H' ~\vdash~\exists~b,  B * H'\tag{$v \subseteq a\cup p \cup b \cup q$}\\
    \Leftarrow&~\forall~a~p~b~q, B * H' ~\vdash~\exists~b,  B * H' \tag{def. of $\wand$}\\
    \Leftarrow&~\mathsf{True}
\end{align*}
\end{subequations}
The observation behind the choice of $v$ (inspired by the \textsc{Localize} rule~\cite{wang2019certifying}) is that, if we want to apply $\phi$ in the first step of the above proof, we have to ensure that the free variables (universal quantifiers) of $\phi$ must be a subset of $a$. Thanks to $v$ capturing the variables that could be outside of $a$, we can prove the above property and apply $\phi$.


\textit{Incorporating the pattern and check part.} Consider the following strategy:
\begin{lstlisting}[numbers=left, xleftmargin=2em]
left:   lseg(?p, ?q, ?l)
left:   p != q // or check: infer(p != q)
action: ... // unfold lseg(p, q, l)
\end{lstlisting}
This strategy unfolds \textbf{lseg} predicate only when it can be unfolded (either $p\ne q$ is in the antecedent, or can be inferred from the antecedent by the SMT solver). If this information is omitted, the generated condition will be unprovable.

The solution is a little bit tricky but simple. For every pattern condition $f,$ we will add two \textit{virtual} operations to the front of the action part: \lstinline|erase(f)| immediately followed by \lstinline|add(f)|, to indicate the existence of $f$. For every check $\textbf{infer}(p)$, we introduce a new \textit{virtual} operation $\textbf{assume}(p)$, which does nothing but \textit{assumes} that the pure formula $p$ is valid. These virtual operations will only be added when generating soundness conditions (and thus will not affect the runtime efficiency when performing automation).

Finally, we formally state the reduction soundness theorem:
\begin{theoremstar}[Reduction Soundness] \label{theorem:soundness}
For every strategy $S$ and every well-formed entailment $E$ and $E'$, if $S$ is applicable for transforming $E$ to $E'$, and $\llbracket \phi(S)\rrbracket$ holds, then $\llbracket E' \rrbracket \Rightarrow \llbracket E \rrbracket$ is valid.
\end{theoremstar}

\textit{Dealing with the instantiation action.} We have also proved in Rocq that applying a strategy with instantiation action is always sound, which is unsurprising since the action only chooses an instantiation for an existential variable.

\textit{Remark.} The generated soundness condition can usually be simplified into a more manageable form, and we provide the \lstinline|pre_process| tactic in Rocq to automate this simplification. As an example, the soundness condition of strategy S0 presented in Section~\ref{sec:overview:2} is actually a simplified version. The raw version is:
\begin{align*}
&\mathsf{lseg}(p, q, l_1) ~\vdash~ \mathsf{emp} * (\forall~l_2~l_3,
(\mathsf{listrep}(q, l_3) \land l_2 == \mathsf{app}(l_1, l_3))\wand \mathsf{listrep}(p, l_2))
\end{align*}
\section{Implementation}\label{sec:implementation}

\textit{Implementation workflow.} Figure~\ref{fig:workflow} presents the workflow of our framework. \dsl{}'s interpreter implements the operational semantics described in Section~\ref{sec:strategy:2}. It reads strategies from the strategy library, applies them based on their priority, and produces three artifacts: (1) output entailments, representing the purified results of the input entailments; (2) side conditions, generated whenever a strategy with the check \lstinline|infer(p)| is applied; and (3) reduction traces, which record the sequence of applied strategies, as well as the proof of each side condition. We implemented a custom SMT solver (supporting EUF and LIA theories) to generate Rocq proofs for side conditions. \dsl{}'s proof constructor generates a sequence of tactics (according to the reduction trace), each of which leverages the soundness condition of corresponding strategy and our core reduction soundness theorem to finally prove that the simplified output entailment entails the original input entailment.

\textit{Rocq implementation of reduction soundness theorem.} Given the imperative nature of our strategy actions and the inability to manipulate entailments (Rocq propositions) directly, we employ the technique of \emph{proof by reflection}~\cite{harrison1995metatheory, chlipala2008certified} by defining an abstract syntax tree (deep embedding) for separation logic entailments. This allows us to formalize strategy actions as functions transforming ASTs and proves the core soundness theorem as detailed in Section~\ref{sec:soundness}.

\textit{Implementation language.} We chose to implement \dsl{} in C instead of in a proof assistant like Rocq for two reasons: (1) C's runtime efficiency is important for a fundamental tool like an entailment purifier; and (2) a C implementation facilitates integration with existing annotation verifiers for tasks such as frame inference--in fact, we have already integrated \dsl{} into an in-house verifier to collect the benchmarks used in our evaluation.

\section{Evaluation}\label{sec:evaluation}
With such convenience provided for writing strategies, it's crucial to ensure that this convenience does not come at the cost of practicality, or more specifically, effectiveness, efficiency, and strategy reusability. Hence, our evaluation aims to address the following research questions:
\begin{enumerate}[leftmargin=*, align=left]
    \item[\textbf{RQ1.}] Is our framework \textit{effective}? Can it automatically purify entailments using the strategies written in \dsl{}?
    \item[\textbf{RQ2.}] Is our framework \textit{efficient}? Can it perform automation within a reasonable amount of time?
    \item[\textbf{RQ3.}] Can our framework facilitate \textit{reuse} of existing automation strategies for purifying new entailments?
\end{enumerate}


\textit{Data collection.} To ensure the practical relevance of our dataset, we collected proof obligations generated from the verification of standard linked data structures and the memory module of a microkernel~\cite{openharmony_liteos_m} (using an in-house annotation verifier).

Specifically, the standard benchmarks include: basic operations on singly-linked lists (\texttt{sll\_basic}) and binary search trees (\texttt{bst\_basic}); merge-sort algorithm (\texttt{sll\_merge}); as well as standard and functional queues implemented via singly-linked lists (\texttt{queue\_sll}, \texttt{fqueue\_sll}) and doubly-linked lists (\texttt{queue\_dll}).

For the microkernel memory module, verification centers on two key data structures: (1) \texttt{freeList}, an array of doubly-linked lists managing the headers of free memory blocks; and (2) \texttt{pool}, a doubly-linked list maintaining all memory blocks and their metadata. The verification maintains a critical structural invariant between them: every header in \texttt{freeList} corresponds to a free block in \texttt{pool}, and conversely, every free block in \texttt{pool} is indexed in \texttt{freeList}. Based on this invariant, we verified the functional correctness of \texttt{OsMemAlloc}, which ensures the allocation of adequately sized blocks (if the return value is not null), and \texttt{OsMemFree}, which reclaims memory and re-establishes the invariant.


\begin{table*}[t]
\caption{Evaluation results}
\label{fig:evaluation}
\begin{center}
    \begin{tblr}{
      width = \linewidth,
      colspec = {X[2, c] | X[2.2, c] | X[6, l] | X[2, c]},
      row{1} = {font=\bfseries}
    }
\toprule
Example & Entailments & \SetCell[r=1]{c} Strategies & Time (s) \\
\midrule
sll\_basic & 48/49 & \textcolor{violet}{common(13)} + \textcolor{blue}{sll(21)} &  0.050 \\
sll\_merge & 42/44 & \textcolor{violet}{common(13)} + \textcolor{blue}{sll(21)} & 0.052 \\
queue\_sll & 10/10 & \textcolor{violet}{common(13)} + \textcolor{blue}{sll(21)} + 2 & 0.021 \\
fqueue\_sll &  20/20  & \textcolor{violet}{common(13)} + \textcolor{blue}{sll(21)} + 2 & 0.019 \\
\midrule
queue\_dll & 14/14 & \textcolor{violet}{common(13)} + dll(13) & 0.024 \\
\midrule
bst\_basic & 6/8 & \textcolor{violet}{common(13)} + bst(12) & 0.013 \\
\midrule
memory & 79/84 & \textcolor{violet}{common(13)} + memory(37) & 0.153 \\
\midrule[0.8pt]
Total & 219/229 & \SetCell[r=1]{c} 98 & 0.332 \\
\bottomrule
\end{tblr}
  \end{center}
\vspace{-1em}
\end{table*}

\textit{Evaluation results.} Table~\ref{fig:evaluation} presents the results of these experiments. In the ``Entailments'' column, $n/m$ indicates that $n$ of the $m$ total entailments were successfully purified. The ``Strategies'' column shows the number of strategies used for purification, where $\mathrm{A}(n)$ indicates a strategy library named $\mathrm{A}$ with $n$ strategies, and a single number $m$ indicates $m$ temporary strategies for unfolding predicates.

The results demonstrate that our framework successfully purifies 95.6\% (219 out of 229) of the entailments, showcasing a high degree of automation. We further checked that all resulting first-order logic entailments are provable, ensuring that our framework does not turn any provable entailment into unprovable. Therefore, we can say \textbf{yes} to \textbf{RQ1}.

Our evaluation also shows that the framework is highly efficient. It took approximately 0.33 seconds to automate all entailments, with an average time of about 1.5 milliseconds per entailment. Thus, we can say \textbf{yes} to \textbf{RQ2}.

Furthermore, the results indicate that our framework supports strategy reuse. For instance, the \textcolor{violet}{common} strategy library, which contains strategies for points-to predicates, is reused across all examples. Moreover, strategies are reusable within the same problem domain. The \textcolor{blue}{sll} strategy library, with strategies for singly-linked list predicates, can be reused for verifying different list operations (e.g., ``length'', ``reverse'', and ``append''). It is also reused in verification of the merge-sort algorithm and queue operations, since their predicates are defined using singly-linked list predicates, and a few unfolding strategies suffice to enable this reuse. Given the above facts, we can say \textbf{yes} to \textbf{RQ3}.
\section{Related Work} \label{sec:related}

In this section, we first discuss related work on the automation of separation logic entailments, which are classified into two categories: rule-based methods and algorithm-based methods. We then discuss related work on other proof automation strategy languages.

\paragraph{Rule-based methods.} SLEEK~\cite{chin2012automated} is an automated separation logic entailment solver that supports extensible automation through user-defined rules. Diaframe \cite{mulder2022diaframe, mulder2023proof} is a framework for the automatic verification of concurrent programs. It introduces a specialized hint format, which is essentially a separation logic rule with limited pattern-matching support (matched formulas must be erased), and is capable of verifying fine-grained concurrent programs mostly automatically.
The primary distinction between these works and \dsl{} lies in the language design. While they rely on rules to perform automation, we offer users a strategy language which eases the process of describing strategies, and our soundness algorithm will automatically generate the appropriate rule. At a technical level, SLEEK relies on proof search and backtracking while \dsl{} is by design non-backtracking, which gives \dsl{} potential to scale for large, real-world settings. This is due to our language design, where one can write more targeted and conservative strategies, which eliminates the need for backtracking under most scenarios. As for Diaframe, our generated rule format (which contains magic wand) is generally more expressive than their hint format. As a result, there exists some strategies which cannot be expressed using Diaframe's hint format (detailed in Appendix~\ref{appendix:related}).

\paragraph{Algorithm-based methods.} Another line of work involves devising clever algorithms to solve entailments. Berdine et al.~\cite{berdine2006smallfoot} propose a decidable fragment (symbolic-heap with linked list predicates) of separation logic and show that the entailment problem lies in coNP. Cook et al.~\cite{cook2011tractable} improve this result by developing a polynomial-time algorithm for the same fragment. Iosif et al.~\cite{iosif2013tree} restrict the predicates to an expressive class with bounded tree width and prove that the entailment problem is decidable. Piskac et al.~\cite{piskac2013automating, piskac2014automating} define a decidable fragment of first-order logic and solve entailments by translating them into this fragment, then leveraging SMT solvers for resolution. Chu et al.~\cite{chu2015automatic} extend the approach of SLEEK~\cite{chin2012automated} by incorporating mathematical induction, while Ta et al. further generalize this approach to support mutual induction~\cite{ta2016automated} and synthesize helper lemmas during induction proofs~\cite{ta2017automated}.
Compared to these methods, our approach is more extensible, since it is easier to design new strategy libraries than to improve previous algorithms. Essentially, our approach is synergistic with these algorithm-based methods: Entailments can first be simplified by our method, rendering them more amenable to these methods; concurrently, these methods are well-suited to automatically discharge the soundness conditions of our strategies.

\paragraph{Proof automation strategy language.} A wide variety of proof automation strategy languages exist, including Ltac~\cite{delahaye2000tactic, pedrot2019ltac2} and Mtac~\cite{ziliani2013mtac, kaiser2018mtac2} in Rocq, Eisbach~\cite{matichuk2016eisbach} and PSL~\cite{nagashima2017proof} in Isabelle, and metaprogramming~\cite{ebner2017metaprogramming} in Lean. These general-purpose languages operate on a broader scope of higher-order logic and support features such as recursion, nested strategies and term rewriting, which \dsl{} does not support.
The most prominent feature of our approach lies in the more flexible goal manipulation. Whereas tactics only allow for sound rule combinations, \dsl{} offers operations that directly add/erase conjuncts, which can potentially introduce unsoundness. This flexibility aligns well with the nature of separation logic entailment solving, where aligning and eliminating corresponding memory layouts are ubiquitous. Nonetheless, this flexibility does not come at the cost of additional soundness effort: Using other strategy languages also requires proving the corresponding rule (e.g., to perform the strategy discussed in the introduction in other strategy languages, one need to first define and prove the corresponding rule as a lemma, and then apply the lemma to manipulate the goal).
\section{Conclusion}\label{sec:conclusion}
This paper presents \textbf{\dsl{}}, a domain-specific strategy language for purifying separation logic entailments. \dsl{} enables manipulating the entailments by adding and erasing conjuncts, offering users \emph{significant convenience} in defining automation strategies. We propose a systematic approach to ensure the strategy soundness, and implement a \emph{foundationally-sound} system for purifying entailments. Our evaluation demonstrates the practicality of this system, which \emph{successfully purifies 95.6\% (219 out of 229) of the entailments} collected from the verification of a set of real-world C programs. These results demonstrate that a strategy-based method can achieve high effectiveness in verifying real-world programs.

\bibliographystyle{splncs04}
\bibliography{main}

\newpage
\appendix
\section{More Strategy Examples} \label{appendix:strategy-example}

\textit{Instantiation of existential variables.} Existential variables often need to be instantiated during the automation. We may desire the following strategy that performs instantiation together with other operations:
\begin{lstlisting}[numbers=left, xleftmargin=2em]
left: data_at(?p, ?v0)
right: data_at(p, ?v1)
action: left_erase(data_at(p, v0));
        right_erase(data_at(p, v1));
        instantiate(v1 -> v0);
\end{lstlisting}
Even though the strategy syntax won't accept this definition, since the instantiation action cannot be combined with other operations, we can always use the following alternative version:
\begin{lstlisting}[numbers=left, xleftmargin=2em]
left: data_at(?p, ?v0)
right: data_at(p, ?V1)
action: left_erase(data_at(p, v0));
        right_erase(data_at(p, v1));
        right_add(v1 == v0);
\end{lstlisting}
where \texttt{instantiate} is replaced by \texttt{right\_add}, and a subsequent application of the following strategy to exactly mimic the behavior of the original strategy:
\begin{lstlisting}[numbers=left, xleftmargin=2em]
priority: 0
right:  exists x, ?x == ?y
action: instantiate(x -> y);
\end{lstlisting}
It is also worth noting that sometimes only partial information about an existential variable is available. For example, in strategy S0 (in Figure~\ref{fig:overview-example}), it is concluded that \lstinline|l1| is the prefix of \lstinline|l2|, but the remainder of \lstinline|l2| is unknown. This can be handled by introducing a fresh existential variable to represent the unknown part (Line~8 in Figure~\ref{fig:overview-example}) before instantiation. Strategy~S4 in Figure~\ref{fig:strategy-example} exhibits a similar scenario where we know that the existential variable \lstinline|l1| should be equal to \lstinline|l| except for the element at index \lstinline|i|. We thus introduce a new existential variable \lstinline|v| to represent this unknown element, and concludes that \lstinline|l1 == update_nth(i - x, v, l)|.

\textit{Deriving pure facts from spatial predicates.} At the beginning of automation, it is often useful to introduce pure facts into the antecedent to facilitate subsequent strategies. For example, given two distinct points-to predicates $\store{p}{v_0}$ and $\store{q}{v_1}$ in the antecedent, we can introduce $p \ne q$ using the following strategy:
\begin{lstlisting}[numbers=left, xleftmargin=2em]
priority: 0
left:   data_at(?p, ?v0)
        data_at(?q, ?v1)
check:  left_absent(p != q);
action: left_add(p != q);
\end{lstlisting}
The check \lstinline|left_absent(p != q)| ensures the strategy is not repeatedly applied, preventing an infinite loop. Assigning the highest priority to this strategy guarantees that it is executed early, thereby avoiding potential information loss caused by other strategies removing the points-to predicates, making it impossible to introduce the pure fact $p \ne q$ later.

\textit{Doubly-linked lists.} We use the predicate $\mathsf{dllseg}(px, x, py, y, l)$ to represent a doubly-linked list segment from $x$ (inclusive) to $y$ (non-inclusive) with content $l$, where $px$ and $py$ denote the previous nodes of $x$ and $y$, respectively. We sometimes want to access the elements within a doubly-linked list, which requires unfolding the \textsf{dllseg} predicate. Unlike singly-linked lists, which admit only a single unfolding direction, doubly-linked list segments can be unfolded from two directions--either from the front ($x$) or from the back ($py$). Instead of attempting to guess the direction via proof search, \dsl{} offers a simple solution: By inspecting the consequent to check which node is being accessed, as shown by the following two strategies:
\begin{lstlisting}[numbers=left, xleftmargin=2em]
left:   dllseg(?px, ?x, ?py, ?y, ?l)
        x != y
right:  data_at(field_addr(x, data), ?v)
action: ... // unfold from the front
\end{lstlisting}
\begin{lstlisting}[numbers=left, xleftmargin=2em]
left: dllseg(?px, ?x, ?py, ?y, ?l)
      x != y
right:  data_at(field_addr(py, data), ?v)
action: ... // unfold from the back
\end{lstlisting}

\section{A Complete Programming Example} \label{appendix:program-example}
\begin{figure}[t]
\begin{lstlisting}[numbers=left, xleftmargin=2em, mathescape]
/*@ $0 < n \land \mathsf{store\shortus array}(a, 0, n, l_a) * \mathsf{store\shortus array}(b, 0, n, l_b)$ */
int i = 0;
/*@ Invariant
    $\exists~l_b, 0 < n \land 0 \le i \le n \land \mathsf{neg}(i, l_a, l_b) \land{}$
      $\mathsf{store\shortus array}(a, 0, n, l_a) * \mathsf{store\shortus array}(b, 0, n, l_b)$ */
while (i < n) {
    b[i] = -a[i];
    i = i + 1;
}
/*@ $\exists~l_b, 0 < n \land \mathsf{neg}(n, l_a, l_b) \land \mathsf{store\shortus array}(a, 0, n, l_a) * \mathsf{store\shortus array}(b, 0, n, l_b)$ */
\end{lstlisting}
    \caption{A complete programming example}
    \label{fig:complete-program}
\end{figure}

\begin{figure}[t]
    \centering
\begin{minipage}[t]{0.48\linewidth}
        \vspace{0pt}
\begin{lstlisting}[basicstyle=\ttfamily\scriptsize]
// Strategy S3
left: store_array(?p, ?x, ?y, ?l)
right: data_at(p + 4 * ?i, ?v)
check: infer(x <= i);
       infer(i < y);
action:
  left_erase(store_array(p, x, y, l));
  right_erase(data_at(p + 4 * i, v));
  left_add(store_array_hole(p, x, y, i, l));
  right_add(v == nth(i - x, l));
\end{lstlisting}
    \end{minipage}
    \hfill
    \begin{minipage}[t]{0.5\linewidth}
\begin{lstlisting}[basicstyle=\ttfamily\scriptsize]
// Strategy S4
left: store_array_hole(?p, ?x, ?y, ?i, ?l)
right: store_array(p, x, y, ?l1)
check: infer(x <= i);
       infer(i < y);
action:
  left_erase(store_array_hole(p, x, y, i, l));
  right_erase(store_array(p, x, y, l1));
  exist_add(v);
  right_add(data_at(p + 4 * i, v));
  right_add(l1 == update_nth(i - x, v, l));
\end{lstlisting}
    \end{minipage}

    \begin{minipage}[t]{0.48\linewidth}
\begin{lstlisting}[basicstyle=\ttfamily\scriptsize]
// Strategy S5
right : exists x, ?x == ?y
action : instantiate(x -> y);
\end{lstlisting}
    \end{minipage}
    \hfill
    \begin{minipage}[t]{0.5\linewidth}
\begin{lstlisting}[basicstyle=\ttfamily\scriptsize]
// Strategy S6
right : ?x == ?x
action : right_erase(x == x);
\end{lstlisting}
    \end{minipage}

    \begin{minipage}[t]{0.48\linewidth}
\begin{lstlisting}[basicstyle=\ttfamily\scriptsize]
// Strategy S7
left : data_at(?x, ?v0)
right : data_at(x, ?v1)
action :
  left_erase(data_at(x, v0));
  right_erase(data_at(x, v1));
  right_add(v1 == v0);
\end{lstlisting}
    \end{minipage}
    \hfill
    \begin{minipage}[t]{0.5\linewidth}
\begin{lstlisting}[basicstyle=\ttfamily\scriptsize]
// Strategy S8
left : store_array(?p, ?x, ?y, ?l1)
right : store_array(p, x, y, ?l2)
action :
  left_erase(store_array(p, x, y, l1));
  right_erase(store_array(p, x, y, l2));
  right_add(l2 == l1);
\end{lstlisting}
    \end{minipage}
    \caption{A strategy set used for the complete example}
    \label{fig:strategy-more-example-1}
\end{figure}

In this part, we show a complete example (Figure~\ref{fig:complete-program}) and exhibit how strategies in Figure~\ref{fig:strategy-more-example-1} can be employed for both frame inference and entailment purification. The predicate $\mathsf{store\shortus array}$ is defined in Section~\ref{sec:strategy:3}, while $\mathsf{neg}(n, l_1, l_2)$ is a pure predicate defined as follows:
\[
\mathsf{neg}(n, l_1, l_2)\triangleq \forall~i, 0\le i < n \to \mathsf{nth}(i, l_1) = -\mathsf{nth}(i, l_2)
\]

\textit{Frame inference.} In Line~7, we encounter a memory read from $a[i]$ and a memory write to $b[i]$. The symbolic state right before the program point is:
\[
\exists l_b, 0 \le i < n \land \mathsf{neg}(i, l_a, l_b) \land \mathsf{store\shortus array}(a, 0, n, l_a) * \mathsf{store\shortus array}(b, 0, n, l_b)
\]
and we should establish (here the existential variable $l_b$ in the antecedent is lifted):
\begin{align*}
&0 \le i < n \land \mathsf{neg}(i, l_a, l_b) \land \mathsf{store\shortus array}(a, 0, n, l_a) * \mathsf{store\shortus array}(b, 0, n, l_b) \vdash\\
&\quad\exists~v_a~v_b, (a + 4 * i) \mapsto v_a * (b + 4 * i) \mapsto v_b * \mathord{?}Q
\end{align*}
and infer the appropriate frame $Q$.

We can perform the following purification in \dsl{}:
\begin{subequations}
\begin{align*}
&~0 \le i < n \land \mathsf{neg}(i, l_a, l_b) \land \mathsf{store\shortus array}(a, 0, n, l_a) * \mathsf{store\shortus array}(b, 0, n, l_b) \vdash\\
&~\quad\exists~v_a~v_b, (a + 4 * i) \mapsto v_a * (b + 4 * i) \mapsto v_b\displaybreak[0]\\
\leadsto^2 &~0 \le i < n \land \mathsf{neg}(i, l_a, l_b) \land{}\\
&~\quad\mathsf{store\shortus array\shortus hole}(a, 0, n, i, l_a) * \mathsf{store\shortus array\shortus hole}(b, 0, n, i, l_b) \vdash\\
&~\quad\exists~v_a~v_b, v_a = \mathsf{nth}(i - 0, l) \land v_b = \mathsf{nth}(i - 0, l_b) \tag{apply S3 twice}\displaybreak[0]\\
\leadsto^4 &~0 \le i < n \land \mathsf{neg}(i, l_a, l_b) \land{}\\
&~\quad\mathsf{store\shortus array\shortus hole}(a, 0, n, i, l_a) * \mathsf{store\shortus array\shortus hole}(b, 0, n, i, l_b) \vdash\\
&~\quad \mathsf{emp} \tag{apply S5 and S6 twice}
\end{align*}
\end{subequations}
Now that we successfully purified the entailment and get the corresponding instantiations (required for performing memory read), we can directly adopt the antecedent of the purified entailment as the inferred frame $Q$, and the current symbolic state becomes:
\begin{align*}
&\exists~l_b, 0 \le i < n \land \mathsf{neg}(i, l_a, l_b) \land{}\\
&\quad\mathsf{store\shortus array\shortus hole}(a, 0, n, i, l_a) * \mathsf{store\shortus array\shortus hole}(b, 0, n, i, l_b) *\\
&\quad(a + 4 * i)\mapsto \mathsf{nth}(i - 0, l_a) * (b + 4 * i) \mapsto \mathsf{nth}(i - 0, l_b)
\end{align*}
which is well-suited to perform the memory read as well as the memory write.

\textit{Purification of proof obligations.} There are three proof obligations: (1) the symbolic state at the loop entrance (Line 3) entails the loop invariant, (2) the preservation of loop invariant, and (3) the loop invariant entails the post-condition (Line 10).

Purification of proof obligation (1) proceeds as follows:
\begin{align*}
&~0 < n \land i = 0 \land \mathsf{store\shortus array}(a, 0, n, l_a) * \mathsf{store\shortus array}(b, 0, n, l_b)\vdash \\
&~\quad\exists~l_b', 0 < n \land 0 \le i \le n \land \mathsf{neg}(i, l_a, l_b') \land{}\\
&~\quad\quad\mathsf{store\shortus array}(a, 0, n, l_a) * \mathsf{store\shortus array}(b, 0, n, l_b')\\
\leadsto^2&~0 < n \land i = 0 \vdash \exists~l_b', 0 < n \land 0 \le i \le n \land \mathsf{neg}(i, l_a, l_b) \land l_a = l_a \land l_b'=l_b \tag{apply S8 twice}\\
\leadsto^2&~0 < n \land i = 0 \vdash 0 < n \land 0 \le i \le n \land \mathsf{neg}(i, l_a, l_b)\tag{apply S5 and S6 repeatedly}\\
\end{align*}

Purification of proof obligation (2) proceeds as follows:
\begin{align*}
&~0 \le i < n \land \mathsf{neg}(i, l_a, l_b) \land{}\\
&~\quad\mathsf{store\shortus array\shortus hole}(a, 0, n, i, l_a) * \mathsf{store\shortus array\shortus hole}(b, 0, n, i, l_b) *\\
&~\quad(a + 4 * i)\mapsto \mathsf{nth}(i - 0, l_a) * (b + 4 * i) \mapsto -\mathsf{nth}(i - 0, l_a) \vdash \\
&~\quad\exists~l_b', 0 < n \land 0 \le i + 1 \le n \land \mathsf{neg}(i + 1, l_a, l_b') \land{}\\
&~\quad\quad\mathsf{store\shortus array}(a, 0, n, l_a) * \mathsf{store\shortus array}(b, 0, n, l_b')\displaybreak[0]\\
\leadsto^2&~0 \le i < n \land \mathsf{neg}(i, l_a, l_b) \land{}\\
&~\quad(a + 4 * i)\mapsto \mathsf{nth}(i - 0, l_a) * (b + 4 * i) \mapsto -\mathsf{nth}(i - 0, l_a) \vdash \\
&~\quad\exists~l_b'~v_a'~v_b', 0 < n \land 0 \le i + 1 \le n \land \mathsf{neg}(i + 1, l_a, l_b') \land{}\\
&~\quad\quad l_a = \mathsf{update\shortus nth}(i - 0, v_a', l_a) \land l_b' = \mathsf{update\shortus nth}(i - 0, v_b', l_b)\\
&~\quad\quad (a + 4 * i) \mapsto v_a' * (b + 4 * i) \mapsto v_b' \tag{apply S4 twice}\displaybreak[0]\\
\leadsto^2&~0 \le i < n \land \mathsf{neg}(i, l_a, l_b)\vdash\\
&~\quad\exists~l_b'~v_a'~v_b', 0 < n \land 0 \le i + 1 \le n \land \mathsf{neg}(i + 1, l_a, l_b') \land{}\\
&~\quad\quad l_a = \mathsf{update\shortus nth}(i - 0, v_a', l_a) \land l_b' = \mathsf{update\shortus nth}(i - 0, v_b', l_b)\\
&~\quad\quad v_a' = \textsf{nth}(i - 0, l_a) \land v_b' = \mathsf{nth}(i - 0, l_b) \tag{apply S7 twice}\displaybreak[0]\\
\leadsto^{*}&~0 \le i < n \land \mathsf{neg}(i, l_a, l_b)\vdash 0 < n \land 0 \le i + 1 \le n \land{}\\
&~\quad\mathsf{neg}(i + 1, l_a, \mathsf{update\shortus nth}(i - 0, -\mathsf{nth}(i - 0, l_a), l_b)) \land{}\\
&~\quad l_a = \mathsf{update\shortus nth}(i - 0, \mathsf{nth}(i - 0, l_a), l_a) \tag{apply S5 and S6 repeatedly}
\end{align*}

Purification of proof obligation (3) proceeds as follows:
\begin{align*}
     &~0 < n \land 0 \le i \le n \land i \ge n \land \mathsf{neg}(i, l_a, l_b) \land{}\\
     &~\quad \mathsf{store\shortus array}(a, 0, n, l_a) * \mathsf{store\shortus array}(b, 0, n, l_b)\vdash\\
     &~\quad \exists~l_b', 0 < n \land \mathsf{neg}(n, l_a, l_b') \land \mathsf{store\shortus array}(a, 0, n, l_a) * \mathsf{store\shortus array}(b, 0, n, l_b')\\
\leadsto^2&~ 0 < n \land 0 \le i \le n \land i \ge n \land \mathsf{neg}(i, l_a, l_b) \vdash\\
&~\quad\exists~l_b, 0 < n \land \mathsf{neg}(n, l_a, l_b') \land l_a = l_a \land l_b = l_b' \tag{apply S8 twice}\\
\leadsto^{*}&~ 0 < n \land 0 \le i \le n \land i \ge n \land \mathsf{neg}(i, l_a, l_b) \vdash 0 < n \land \mathsf{neg}(n, l_a, l_b)\tag{apply S5 and S6 repeatedly}
\end{align*}
\section{Formal Process for Generating Soundness Condition}\label{appendix:soundness}
We use the quadruple $(r, \vec{q}, \vec{c}, \vec{o})$ to represent the strategy $S$. The process proceeds as follows:
\begin{enumerate}
    \item For each left pattern condition $f$ in $\vec{q}$, we prepend the operations \lstinline|left_erase(f)| and \lstinline|left_add(f)| to $\vec{o}$, and perform an analogous procedure for the right pattern conditions, yielding $\vec{o}_1$.
    \item For every check condition \lstinline|infer(p)| in $\vec{c}$, we add an operation \lstinline|assume(p)| to the front of $\vec{o}_1$, resulting in $\vec{o}_2$.
    \item Compute a six-tuple $(vl_{\forall}, sc, l_{-}, l_{+}, r_{+}, r_{-})$ based on $\vec{o}_2$, where:
    \begin{itemize}
        \item \(vl_{\forall}\) collects the variables in the operation \lstinline|forall_add(x)|;
        \item \(sc\) collects the assumed formulas in the operation \lstinline|assume(p)|;
        \item \(l_{-}\) collects the formulas that are removed (but \textit{not previously added}) in the operation \lstinline|left_erase(f)|;
        \item \(l_{+}\) collects the formulas added in the operation
        \lstinline|left_add(f)|;
        \item \(r_{-}\) and \(r_{+}\) are defined similarly for the right-hand side of the entailment.
    \end{itemize}
    \item Compute the set of quantifiers that \textit{only appears in} $r_{+}$ and $r_{-}$ (instead of in $vl_{\forall}, sc, l_{+}$ and $l_{-}$) as $v$.
    \item Generate $\phi(S)$ as the soundness condition:
    $$
    \phi(S)\triangleq sc \land l_{-} ~\vdash~ \exists~vl_{\forall}, l_{+} * (\forall~v, r_{+} \wand r_{-})
    $$
\end{enumerate}

\section{Missing Details in Related Work} \label{appendix:related}
Diaframe's hint format is as follows (here we restrict our discussion to the sequential setting, while their original version is capable of handling the concurrent setting):
\begin{align*}
&H * [\vec{y}; L] \vdash \vec{x}; A * [U] \triangleq \\
&\quad \forall~\vec{y}.H * L \vdash \exists~\vec{x}. A * U
\end{align*}
This hint simplifies an entailment when $H$ matches the antecedent and $A$ matches the consequent. The simplification proceeds via the following inference rule, transforming the goal $E_1$ into $E_2$:
$$
\inference[]{H * [\vec{y}; L] \vdash \vec{x}; A * [U]\\E_2\triangleq \Delta\vdash \exists~\vec{y}. L * (\forall~\vec{x}. U \wand G)}{E_1\triangleq \Delta * H \vdash \exists~\vec{x}. A * G}
$$
Crucially, this inference rule removes the matched conjuncts $H$ and $A$ from the entailment, which limits the expressiveness of strategies definable in Diaframe's format. Specifically, consider the following strategy that conservatively eliminates the \textsf{lseg} predicate:
\begin{lstlisting}
left: lseg(?p, ?q, ?l1)
right: lseg(p, q, ?l2)
left: data_at(field_addr(q, next), ?v1)
right: data_at(field_addr(q, next), ?v2)
action: left_erase(lseg(p, q, l1));
        right_erase(lseg(p, q, l2));
        right_add(l2 == l1);
\end{lstlisting}
As discussed in Section~\ref{sec:overview:1}, where we use an extra \textsf{listrep} predicate to exclude the potential cycle within the \text{lseg} predicate, here shows a similar situation where two extra points-to facts are required to do the same thing. This strategy cannot be expressed in Diaframe's hint format. To ensure the conservative application, the points-to facts must be included in $H$ and $A$, which, however, will be removed after the hint application. Here we indent to preserve these facts, as they may be necessary pattern conditions for subsequent strategies.

One might turn to a workaround where we first remove these points-to facts and then add them back through $L$ and $U$, which will result in the following counter-intuitive hint:
\begin{align*}
&\mathsf{lseg}(p, q, l_1) * \texttt{\&}(q\texttt{.next})\mapsto v_1 * [v_2; \texttt{\&}(q\texttt{.next})\mapsto v_2] \vdash \\
&\quad l_2; \mathsf{lseg}(p, q, l_2) * \texttt{\&}(q\texttt{.next})\mapsto v_2 * [\texttt{\&}(q\texttt{.next})\mapsto v_1]
\end{align*}
which is theoretically valid since the antecedent is vacuously true, its application, however, does not yield the expected result:
\begin{align*}
&\mathsf{lseg}(p, q, l_1) * \texttt{\&}(q\texttt{.next})\mapsto v_1 \vdash \\
&\quad \exists~l_2~v_2, \mathsf{lseg}(p, q, l_2) * \texttt{\&}(q\texttt{.next})\mapsto v_2\\
\leadsto~& \mathsf{emp} \vdash \exists~v_1, \texttt{\&}(q\texttt{.next})\mapsto v_1 * (\forall ~v_2, \texttt{\&}(q\texttt{.next})\mapsto v_2 \wand \mathsf{emp})
\end{align*}
The original provable entailment becomes unprovable after the hint application, as it requires to split a points-to fact from an empty heap.

The fundamental reason \dsl{} is capable of handling such strategies is that our generated soundness condition takes the general form $A \vdash B * (C \wand D)$. This form is strictly more expressive than Diaframe's hint format. While this design choice admittedly imposes a greater proof burden, the resulting soundness conditions can typically be automatically simplified to a form without magic wand. As noted in Section~\ref{sec:soundness}, we provide a Rocq tactic, \lstinline|pre_process|, to automate this simplification.











\end{document}